\documentclass[a4paper,12pt]{article}
\usepackage{graphicx}
\usepackage[round,numbers,sort&compress]{natbib} 
\usepackage{siunitx}
\usepackage{amsmath} 
\usepackage{amssymb}
\usepackage{gensymb} 
\usepackage{subcaption}
\usepackage{xcolor}
\usepackage{indentfirst}
\usepackage{tikz}
\usepackage[font=scriptsize,labelfont=scriptsize,textfont=scriptsize]{caption} 
\usepackage{array} 
\usepackage{booktabs} 
\usepackage[auth-sc,affil-it]{authblk} 
\usepackage{setspace} 
\usepackage{url} 
\usepackage{soul} 
\usepackage{lineno} 
\usepackage{draftwatermark} 
\SetWatermarkText{Accepted: Pharmaceutical Research (2021)}
\SetWatermarkScale{1}
\SetWatermarkAngle{0}
\SetWatermarkVerCenter{0.93\paperheight}

\definecolor{plt.red}{RGB}{214, 39, 40}
\definecolor{plt.blue}{RGB}{31, 119, 180}
\definecolor{plt.green}{RGB}{44, 160, 44}
\definecolor{plt.black}{RGB}{0, 0, 0}

\newcommand{\blueline}{\raisebox{2.0pt}{\tikz{\draw[ultra thick,plt.blue](-0.4,0) -- (0.4,0)}}}

\newcommand{\bluemarkline}{\raisebox{-0.5pt}{\tikz{\draw[thick,plt.blue](-0.4,0) -- (0.4,0); \draw[plt.blue,very thick](0,0)circle(3pt)}}}

\newcommand{\bluebrokenline}{\raisebox{2.0pt}{\tikz{\draw[dashed,ultra thick,plt.blue](-0.4,0) -- (0.4,0)}}}

\newcommand{\greenline}{\raisebox{2.0pt}{\tikz{\draw[ultra thick,plt.green](-0.4,0) -- (0.4,0)}}}

\newcommand{\greenmarkline}{\raisebox{-0.5pt}{\tikz{\draw[thick,plt.green](-0.4,0) -- (0.4,0); \draw[plt.green,very thick](0,0)circle(3pt)}}}

\newcommand{\greenbrokenline}{\raisebox{2.0pt}{\tikz{\draw[dashed, ultra thick,plt.green](-0.4,0) -- (0.4,0)}}}

\newcommand{\redline}{\raisebox{2.0pt}{\tikz{\draw[ultra thick,plt.red](-0.4,0) -- (0.4,0)}}}

\newcommand{\redmarkline}{\raisebox{-0.5pt}{\tikz{\draw[thick,plt.red](-0.4,0) -- (0.4,0); \draw[plt.red,very thick](0,0)circle(3pt)}}}

\newcommand{\redbrokenline}{\raisebox{2.0pt}{\tikz{\draw[dashed, ultra thick,plt.red](-0.4,0) -- (0.4,0)}}}

\newcommand{\blackcircle}{\raisebox{-0.5pt}{\tikz{ \draw[plt.black,very thick](0,0)circle(2.5pt)}}}

\newcommand{\nog}{\textit{no-grid}}
\newcommand{\eng}{\textit{entry-grid}}
\newcommand{\exg}{\textit{exit-grid}}
\newcommand{\invit}{\textit{in-vitro}}

\title{On the Use of Computational Fluid Dynamics (CFD) Modelling to Design Improved Dry Powder Inhalers} 
\author[1]{David F Fletcher}
\author[2*]{Vishal Chaugule}
\author[3]{\authorcr Larissa Gomes dos Reis}
\author[3]{Paul M Young}
\author[3]{\authorcr Daniela Traini}
\author[2]{Julio Soria}
\affil[1]{\footnotesize	{School of Chemical and Biomolecular Engineering, The University of Sydney, Sydney, Australia}}
\affil[2]{\footnotesize	{Laboratory for Turbulence Research in Aerospace and Combustion (LTRAC), Department of Mechanical and Aerospace Engineering, Monash University, Clayton Campus, Melbourne, Australia}}
\affil[3]{\footnotesize	{Respiratory Technology, Woolcock Institute of Medical Research and Discipline of Pharmacology, Faculty of Medicine and Health, The University of
Sydney, Sydney, Australia}}
\affil[*]{Corresponding author: \underline{vishal.chaugule@monash.edu}}
\date{}

\begin{document} 
    \maketitle 

\begin{abstract}
\noindent \textbf{Purpose}\\
Computational Fluid Dynamics (CFD) simulations are performed to investigate the impact of adding a grid to a two-inlet dry powder inhaler (DPI). The purpose of the paper is to show the importance of the correct choice of closure model and modeling approach, as well as to perform validation against particle dispersion data obtained from {\invit} studies and flow velocity data obtained from particle image velocimetry (PIV) experiments.\\
\textbf{Methods}\\
 CFD simulations are performed using the Ansys Fluent 2020R1 software package. Two RANS turbulence models (realisable $k$-$\epsilon$ and $k$-$\omega$ SST) and the Stress Blended Eddy Simulation (SBES) models are considered. Lagrangian particle tracking for both carrier and fine particles is also performed.\\
\textbf{Results}\\
 Excellent comparison with the PIV data is found for the SBES approach and the particle tracking data are consistent with the dispersion results, given the simplicity of the assumptions made.\\
\textbf{Conclusions}\\
This work shows the importance of selecting the correct turbulence modelling approach and boundary conditions to obtain good agreement with PIV data for the flow-field exiting the device. With this validated, the model can be used with much higher confidence to explore the fluid and particle dynamics within the device.  
\end{abstract}


\section*{Abbreviations}
\noindent
API - Active Pharmaceutical Ingredients\\
CC - Curvature Correction\\
CFD - Computational Fluid Dynamics\\
DPI - Dry Powder Inhaler\\
DPM - Discrete Phase Model\\
FPF - Fine Particle Fraction\\
LDV - Laser Doppler Velocimetry\\
LES - Large Eddy Simulation\\
LRN - Low Reynolds Number\\
NSE - Navier-Stokes Equations\\
PIV - Particle Image Velocimetry\\
RANS - Reynolds-Averaged Navier-Stokes\\
SBES - Stress Blended Eddy Simulation\\
SRS - Scale-Resolving Simulation\\
SST - Shear Stress Transport\\
URANS - Unsteady Reynolds-Averaged Navier-Stokes\\
WALE - Wall-Adapting Local Eddy-viscosity\\

\newpage
    \section{Introduction}
Pharmaceutical aerosol generated through a dry powder inhaler (DPI) is a multi-phase flow comprising a continuous phase (air) and a disperse phase (particles), which contains the active pharmaceutical ingredients (API). During aerosolization, there is an interaction between the two phases - the air flow contributes to the dispersion and deposition of the particles, and the presence and motion of particles modulates the air flow-field.  The transition of local flow from laminar to turbulent and the high volume fraction of particles near the release point, relative to the fluid  volume in a DPI, leads to complex particle-flow interactions. In addition, these particle-flow interactions are symbiotic with the device design, the inhalation flow, and the formulation and properties of the drug which further increases its complexity. Experimental investigations of these phenomena have significant practical challenges, thus computational modelling of the fluid flow and particle dynamics has been  performed to study these processes and optimize device delivery \cite{Wong2012, Islam2012, Sommerfeld2019}. 

The modelling of the continuous phase of a DPI has been performed using Computational Fluid Dynamics (CFD), which has traditionally involved solving the Reynolds-Averaged Navier-Stokes (RANS) equations numerically, with suitable turbulence closure models. These equations are time-averaged forms of the governing continuity and momentum equations (Navier-Stokes equations (NSE)), and the turbulence model serves to close this system of mean-flow equations. However, time-averaging leads to a loss of information and some turbulence models have limitations in accurately modelling turbulent swirling flows that are inherent in a DPI \cite{Yang2012}. These issues can be mitigated by using Large Eddy Simulation (LES), that solves the filtered NSE and can resolve large-scale turbulence eddies and detailed flow structures, depending on the applied local filter width. LES has been shown to provide more high-fidelity information of the flow field compared with RANS, but it has not been widely used for DPI modelling because of the higher computational requirements, especially if it is applied in boundary layers \cite{Milenkovic2013}.

One of the earliest CFD studies on DPIs was conducted by Coates et al. \cite{Coates2004-1} in which they studied the flow-field and particle trajectories in the Aerolizer\textsuperscript{\tiny\textregistered} DPI for different design parameters of the inhaler mouthpiece and grid. The flow-field was simulated using the RANS approach with the $k$-$\omega$ Shear Stress Transport (SST) turbulence model \cite{Menter1994} and with particles tracked using a Lagrangian approach. Flow field validation was carried out by comparing the simulation results with laser doppler velocimetry (LDV) data at the exit of the device. An increase in the size of the grid openings reduced the flow straightening effect, and also the turbulence intensity, just downstream of the grid. Consequently, particle collisions with the grid also decreased, but led to an increase in particle-wall collisions in the mouthpiece. This balancing effect, of lower turbulence intensity and particle-grid collisions with higher particle-wall collisions in the mouthpiece, was found to result in similar values of fine particle fraction (FPF) for these design changes.

In a follow-up study on the effect of flow rates on DPI performance \cite{Coates2005}, they reported the expected increase of turbulence intensity, integral scale strain rates and particle-wall collisions with an increase in air flow rates. This led to an improvement in powder de-agglomeration and thus its dispersion in the flow, but only up to a flow rate of 65 l/min. A later study by Coates et al. \cite{Coates2006-2} on the effect of tangential inlet size on the inhaler flow-field showed that a reduction of inlet area size resulted in higher turbulence intensities and velocity of particle-wall collisions in the region just downstream of the inlets. 

A RANS approach using the $k$-$\omega$ SST turbulence model was used by Donovan et al. \cite{Donovan2012} to study the flow-field and particle trajectories in the Aerolizer\textsuperscript{\tiny\textregistered} and Handihaler\textsuperscript{\tiny\textregistered} DPI geometries. The particles were modelled using a Stokesian drag law with non-spherical corrections to account for particle shape effects. The swirling flow in the Aerolizer\textsuperscript{\tiny\textregistered} intensified particle-wall collisions, which lead to an improvement in drug detachment, whereas the absence of swirling flow in the Handihaler\textsuperscript{\tiny\textregistered} lead to fewer particle collisions with the inhaler wall, and thus lower aerosol performance. It was also shown that increasing the mean particle diameter increased the number of particle-wall collisions due to the increased Stokes number leading to more  ballistic trajectories.

The application of RANS with various models for turbulent flow (standard $k$-$\epsilon$, RNG $k$-$\epsilon$ and k-$\omega$ SST) was used by Milenkovic et al. \cite{Milenkovic2013} to model the flow in a Turbuhaler\textsuperscript{\tiny\textregistered} DPI geometry. They also used LES, but for only a single parametric case, which was then compared with the RANS solutions.  The LES generated radial and tangential flows within the device showed enhanced presence of eddies and secondary flow structures that were most similar to those obtained with the $k$-$\omega$ SST model. 
In a later study, Milenkovic et al. \cite{Milenkovic2014} modelled the dynamic flow in the same DPI geometry instead of a steady flow. This dynamic flow comprised an initial rapid increase of flow rate that gradually plateaued to a steady flow rate, and was simulated by imposing dynamic outlet pressures. They showed that the normalised dynamic flow-field velocities were similar for peak inspiratory flow rates (PIFR) of 30, 50 and 70 l/min. 

A Lagrangian approach with one-way coupling was used by Sommerfeld and Schmalfu{\ss} \cite{Sommerfeld2016} to determine the fluid stresses experienced by the carrier particles along their path through a DPI. The RANS equations with the $k$-$\omega$ SST  turbulence model were solved for steady flow through the inhaler. Their results indicated that wall collisions largely prevailed in particle motion, wherein de-agglomeration of drug powder mainly occurred due to wall impacts in the swirl chamber and with the grid placed just after it. The wall-collision frequency of the particles was found to increase with particle size due to their increased inertia, but this reduced their wall-impact velocities. 

Longest et al. \cite{Longest2013} performed CFD simulations using the low Reynolds number (LRN) $k$-$\omega$ turbulence model and employed a Lagrangian particle tracking algorithm to predict individual particle trajectories and determine particle interaction with the mean turbulent flow-field.  Six different inhaler designs were studied and they explored both turbulence and impaction as potential particle break-up mechanisms. It was found that turbulence was the primary de-aggregation mechanism for carrier-free particles, with high turbulence kinetic energy, long exposure time, and small characteristic eddy length scales. However, in a later study by Longest and Farkas \cite{Longest2019}, on powder dispersion in a dose aerosolization and containment unit, they found an undesirable increase in aerodynamic diameter when flow turbulence was increased.

It is important to keep in mind that CFD simulations can only be used with confidence once they have been validated. It is for this reason that we are employing three complementary methods in our current investigation of the impact of inhaler design on performance. CFD can provide information on the flow field  and particle behaviour both inside and outside of the inhaler, however there are many uncertainties pertaining to turbulence modelling and the dynamics and break-up of particle agglomerates.  Particle image velocimetry (PIV) studies provide high quality data on the flow field outside of the device. Finally, {\invit} studies provide a means of studying device performance for a powder formulation and the interaction of the inhaled particle cloud with the respiratory tract. Ultimately, 
models should reliably determine particle deposition inside the device as this in turn affects the determination of emitted FPF from simulations. The size, distribution and velocity of aerosol particles upon exiting the DPI mouthpiece govern their motion and deposition in the respiratory tract, which is of utmost importance in assessing the performance of the DPI. 

\begin{figure}[!h]
\centering      
    \begin{subfigure}{0.3\textwidth}
	\centering
	\includegraphics[width=0.6\textwidth]{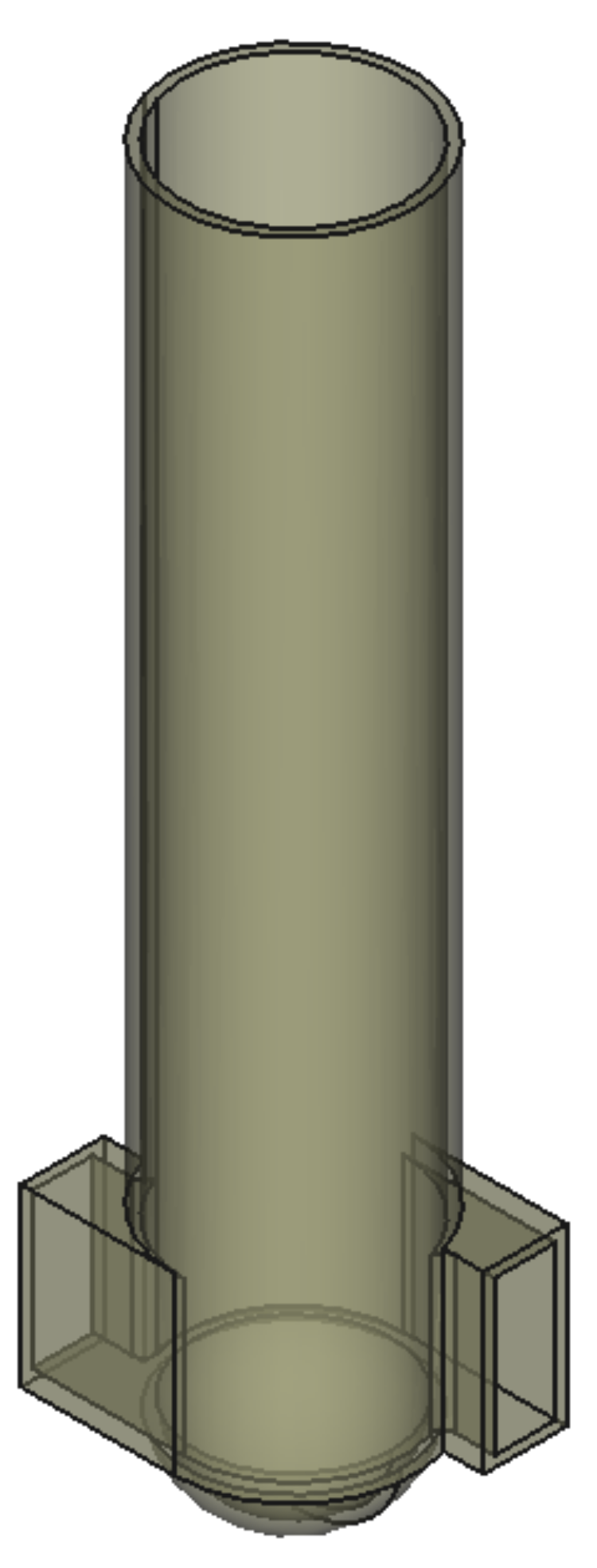} 
	\caption{\nog}
	\end{subfigure} 
	\hspace{0.3em}
	\begin{subfigure}{0.3\textwidth}
	\centering
	\includegraphics[width=0.6\textwidth]{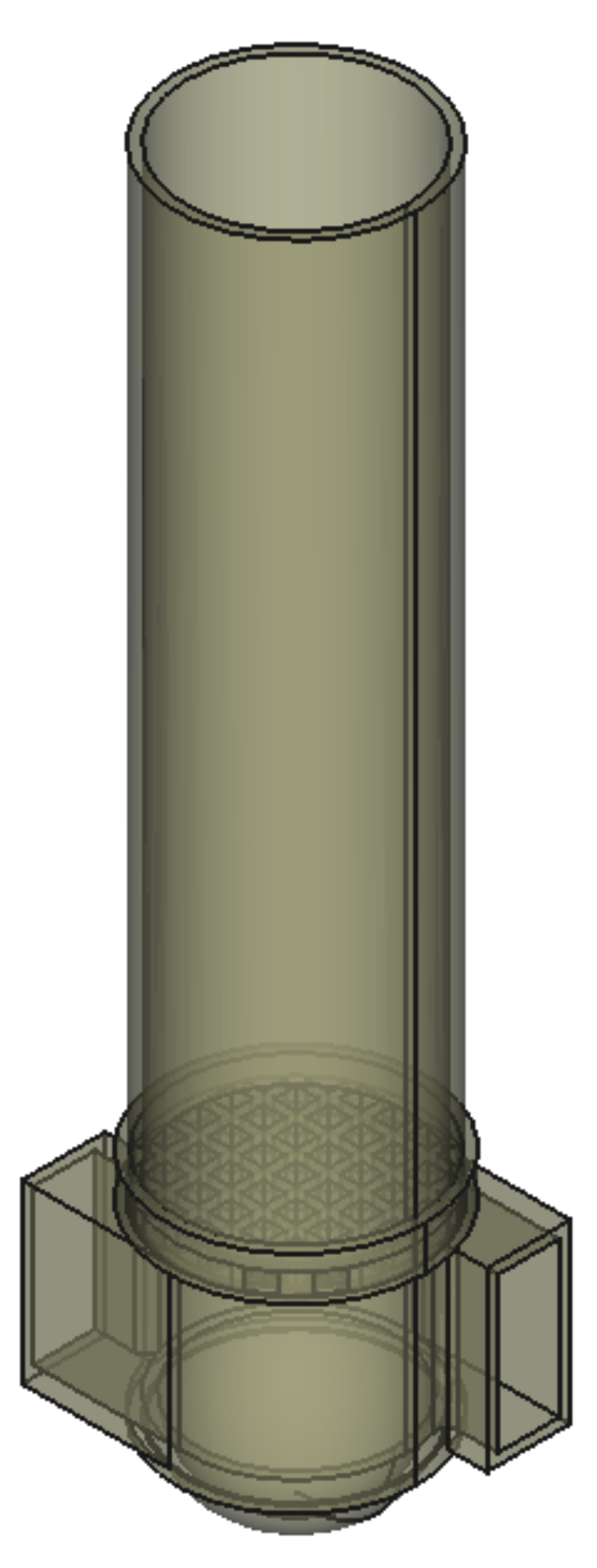} 
	\caption{\eng}
	\end{subfigure} 
	\hspace{0.3em}
	\begin{subfigure}{0.3\textwidth}
	\centering
	\includegraphics[width=0.6\textwidth]{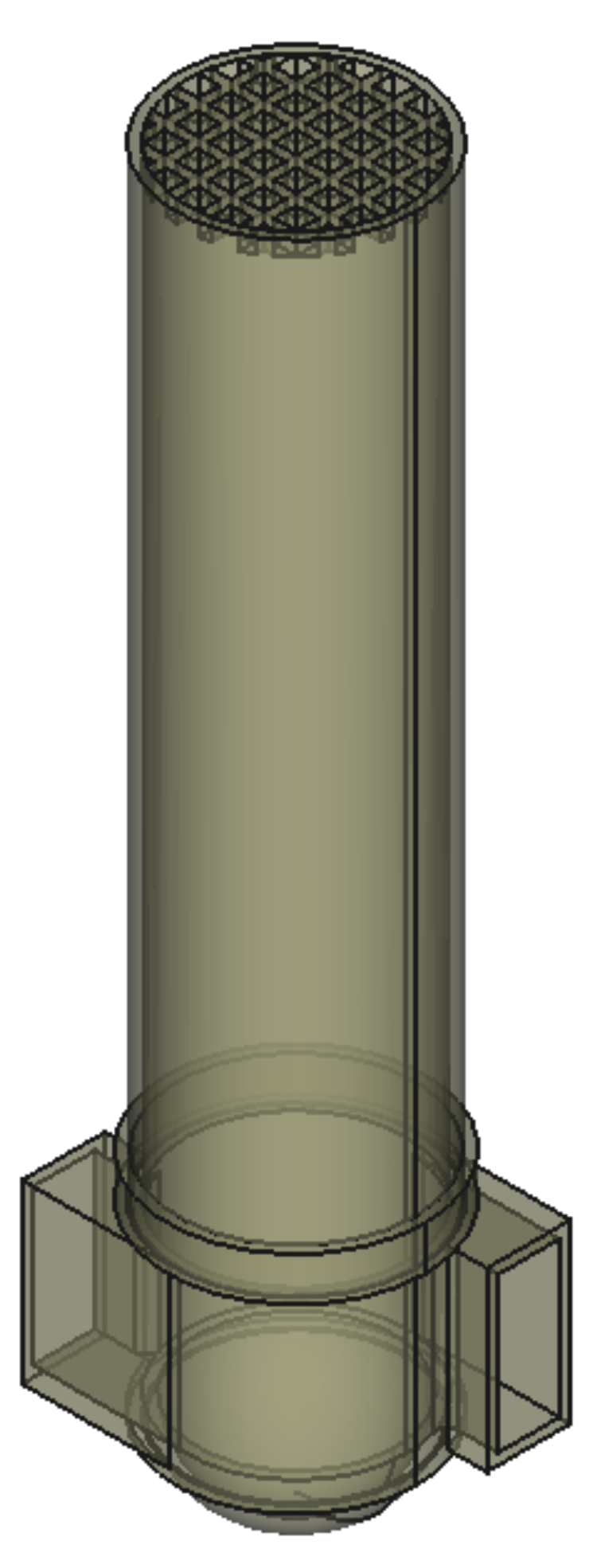} 
	\caption{\exg}
	\end{subfigure} 
\caption{\sl DPI device models examined in this study}
\label{fig:DPI models}
\end{figure}

In a previous study {\cite{dosReis2020}} we presented both PIV data and {\invit} studies for four different inhalers having two tangential inlets, six tangential inlets, two inlets with an inlet grid and two inlets with an exit grid. Given that the two and six inlet cases showed very similar results, in this paper we present a CFD study of the two inlet cases and compare our results with both the {\invit} and PIV data. The inhaler geometries studied here are shown in Figure \ref{fig:DPI models}.

\section{Materials and Methods}

\subsection{PIV Setup}\label{pivsetup}
\begin{figure}[h!]
	\centering
	\includegraphics[scale=0.6]{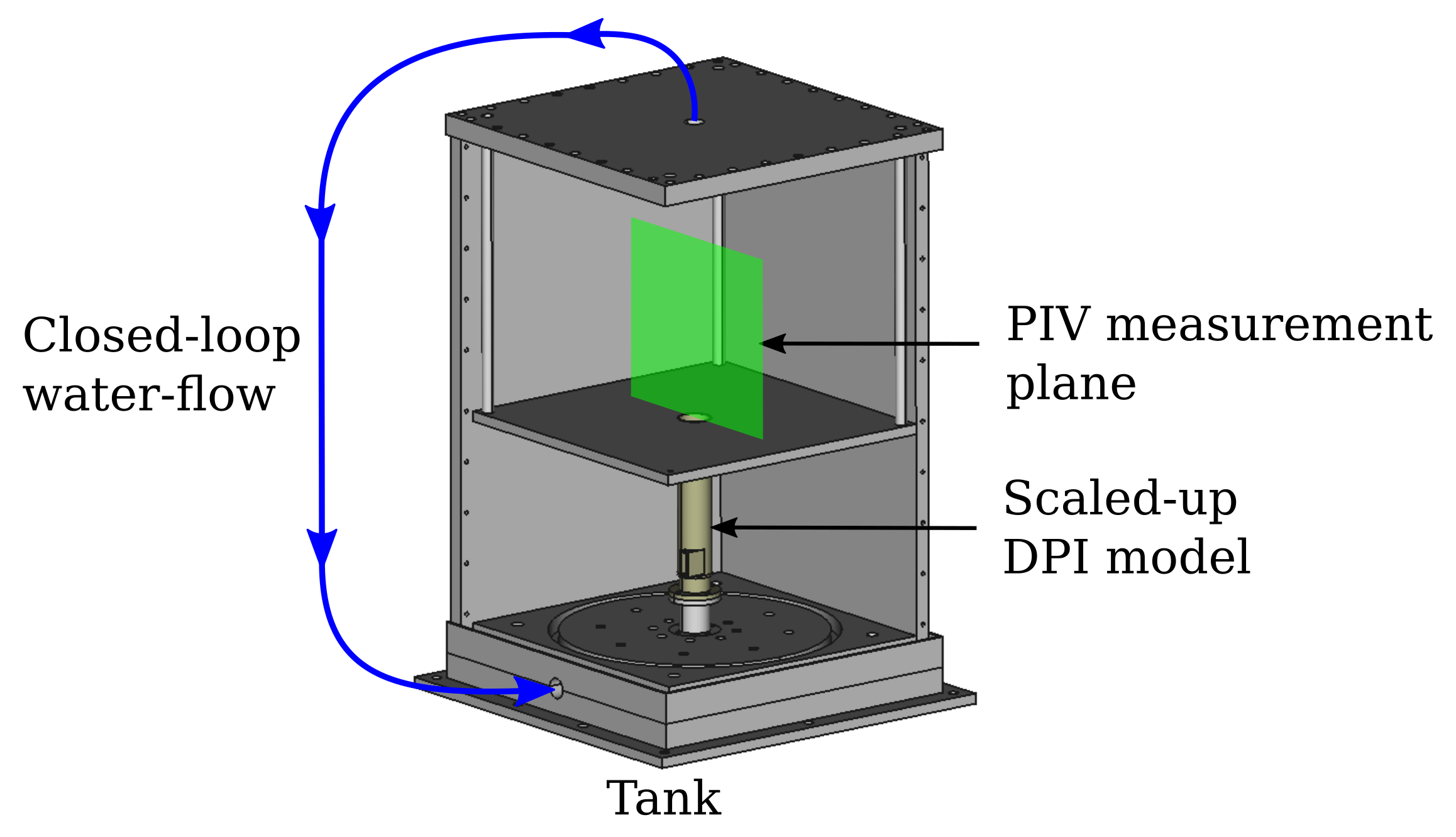} 
	\caption{PIV experimental setup}
\label{fig:PIV-setup}
\end{figure}

The PIV experimental setup, which the CFD model geometry replicates, is shown in Fig. {\ref{fig:PIV-setup}}. The DPI device models used in the PIV experiments were geometrically scaled-up three times to that of the original models shown in Fig. {\ref{fig:DPI models}}. Each model was placed in a tank with a closed-loop water flow system, wherein a steady water flow-rate was maintained through the model to attain a Reynolds number of $\approx$ 8400. The Reynolds number is defined based on the average flow velocity at the DPI mouthpiece exit and the mouthpiece exit inner-diameter. Two component-two dimensional (2C-2D) PIV measurements were performed in a longitudinal plane outside the DPI mouthpiece exit, within a downstream distance of four jet diameters. The detailed description of the PIV apparatus, methodology, and associated measurement uncertainties is provided in Gomes dos Reis et al. {\cite{dosReis2020}}.

\subsection{CFD Modelling Approach}

In all cases time-dependent simulations were performed as, not unexpectedly, convergence to a steady flow could not be achieved. Therefore, simulations were started in steady mode to establish an initial flow field and then time-dependent simulations were performed. Once these had established realistic flow fields, transient statistics were evaluated to enable the mean flow velocities and the Reynolds stresses to be obtained for comparison with the PIV experimental data. All simulations were performed using Ansys\textsuperscript{\tiny\textregistered} Fluent 2020R1 \cite{Ansys2020} and were run in double precision to eliminate rounding error. 

\subsubsection{Turbulence Modelling}
Based on the above literature review it was decided to investigate three different turbulence modelling approaches. The realisable $k$-$\epsilon$ \cite{Shih1995} and the $k$-$\omega$ SST \cite {Menter1994} models were chosen as being representative of the unsteady-RANS (URANS) modelling approaches. It is clear that the $k$-$\omega$ SST model is the most widely used, however a $k$-$\epsilon$ model was also included as this approach is widely used in internal flow simulations. It is well-known that these two-equation models do not capture swirling flow correctly, so both were solved with a Curvature Correction (CC) term included \cite{Smirnov2009}, as it has been shown to correctly capture the swirl profile in cyclones \cite{Alahmadi2016}. Whilst Reynolds stress models can in theory provide good solutions for swirling flows they are renowned for being numerically stiff and hard to solve, so they were not investigated in this study.

In order to investigate the impact of using a Scale-Resolving Simulation (SRS) approach, simulations were made using the Stress Blended Eddy Simulation (SBES) approach \cite{Menter2018} as this takes advantage of the best aspects of the RANS and LES approaches. In the near wall region, where the flow is attached and LES simulations are prohibitively expensive, the $k$-$\omega$ SST model  provides the eddy viscosity. Away from the wall, in regions where the mesh is sufficiently fine, the model blends the eddy viscosity with that from an LES modelling approach. The subgrid-scale closure of the Wall-Adapting Local Eddy-viscosity (WALE) model \cite{Nicoud1999} was used.

In all cases the computational mesh was constructed so that there were sufficient inflation layers adjacent to the inhaler walls that the $y\textsuperscript{+}$ values were low enough for the flow to be resolved up to the wall in the $k$-$\omega$ models. Care was taken to ensure that the transition to SRS occurred where expected and that in this case the unresolved turbulence led to an eddy viscosity consistent with the LES approach. A recent study that highlights the best practices and checks to be performed can be consulted for more detail \cite{Brown2020}. 

\subsubsection{Particle Modelling} \label{particle_model}
  
Once the flow was established, the  Discrete Phase Model (DPM) was used to perform time-dependent particle tracking in the time-dependent flow for the SBES simulations, assuming a drag model appropriate for smooth spheres. The simulations were performed for a low particle loading using one-way coupling  as the current work compares the flow field with PIV data in which the drug particles are absent. As the large scale turbulence structures are captured in these simulations, no additional turbulent dispersion was added. At the walls, particles were assumed to reflect with  coefficients of restitution of 0.9 in the tangential direction and 0.7 in the normal direction, based on values determined for typical drug formulations {\cite{Bharadwaj2010}}. User-defined functions were used to capture the number of impacts  and the impact kinetic energy of the particles.    

Two different sets of particle tracking were performed. Firstly, \SI{280}{\micro\metre} diameter particles were released from the spherical end cap of the inhaler (dosing cup) to represent the carrier particles, and their impact behaviour with the wall and grid (if present) was studied. Particle de-agglomeration occurs when carrier particles impact the wall or each other, knocking active drug particle off the carrier particle. Here we investigated the importance of wall impact by recording both the average number of wall impacts and the average impact kinetic energy of the particles. Secondly, \SI{1.24}{\micro\metre} diameter particles were released from an annulus one nozzle diameter upstream of the mouthpiece exit, occupying the outer 20\% of the device mouthpiece to represent the fine particles. This simulation was made to investigate the subsequent dispersion of these particles assuming they had been released from wall impaction and had subsequently travelled along the wall region. In both cases a particle density of \SI{1540}{\kilo\gram\per\cubic\metre} was used, based on that for lactose \cite{Zuurman1994}.
  
\subsubsection{Model Setup}

\begin{figure}[h!]
	\centering
	\begin{subfigure}{0.45\textwidth}
	\centering
	\includegraphics[scale=0.4]{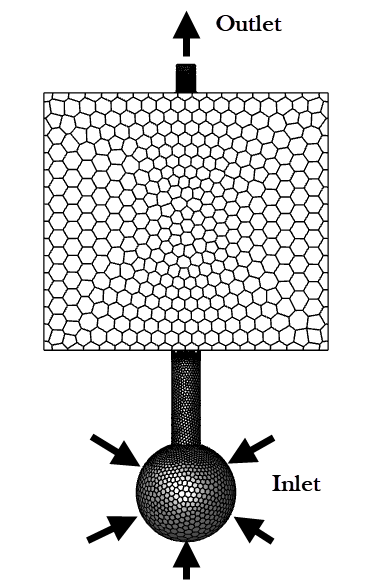} \caption{}
	\end{subfigure} 
	\begin{subfigure}{0.45\textwidth}
	\centering
	\includegraphics[scale=0.3]{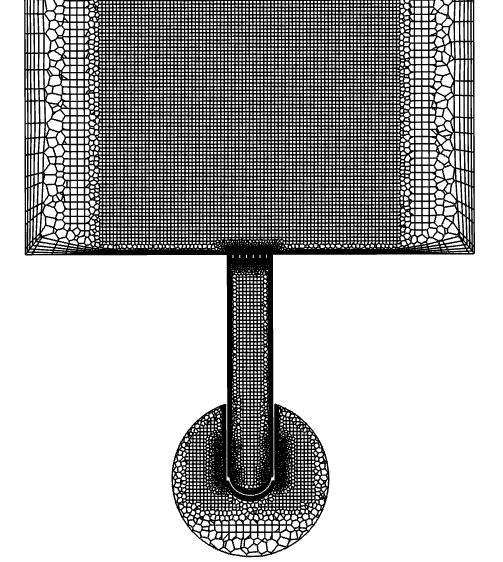} 
	\caption{}
	\end{subfigure} 
\caption{CFD model geometry: (a) Geometry ; (b) Mesh}
\label{fig:Geometry}
\end{figure}

The model geometry was created to mirror that of the PIV experiment, briefly described in Section {\ref{pivsetup}},  but for an incompressible fluid of air, at ambient conditions. The Reynolds number based on the jet diameter $D_a$ was 8400, as used experimentally. The geometry used, showing the external surface mesh, is given in Figure \ref{fig:Geometry}(a). A spherical region of ambient air is modelled around the inlet region, as it was found that applying boundary conditions at the inlets of the inhaler led to an over-constrained flow in that region. The air exiting the device enters a box, just as was used in the PIV experiments, in order provide the same downstream flow domain to allow direct comparison of the jet behaviour with the experimental data. Figure \ref{fig:Geometry}(b) shows a section through the computational mesh for the case with a grid at the exit, showing the poly-hexcore structure used, with hexahedral mesh in the important central regions, connected to inflation mesh at the walls by a layer of polyhedra. Local mesh controls were applied to ensure good resolution where needed. Based on mesh studies, the final mesh comprised $\sim$1 million cells and $\sim$2 million nodes. 

The adequacy of the inflation mesh was checked by examining the wall $y^+$ values. For the SST model  $y^+ < 8$ over all walls, with most of the domain having $y^+ < 3$, meaning that the model was resolving the flow to the wall. For the realisable $k$-$\epsilon$, $ 11 < y^+ < 200$, which was consistent with the use of scalable wall functions.

At the inlet a total pressure of \SI{0}{\pascal} was applied with a 1\%\ turbulence intensity. At the exit the mass flow rate was specified to achieve the required Reynolds number. All walls were treated as being smooth with no slip. 

To solve the equations the coupled solver was applied in time-dependent mode. The very strong swirl meant that a segregated approach was very hard to converge. Gradients were calculated using the Green-Gauss node based method to achieve high accuracy. A second order differencing scheme was used for the pressure, a bounded central differencing scheme for momentum, a second order upwind scheme for the turbulence quantities and a bounded second order implicit scheme for the transient terms.  The solution required the use of small time steps ($\sim$ \SI{5}{\micro\second}) and typically 5 - 8 iterations per timestep.

\section{Results}

\subsection{Effect of Turbulence Model}

\begin{figure}[h!]
	\centering
	\includegraphics[scale=0.5]{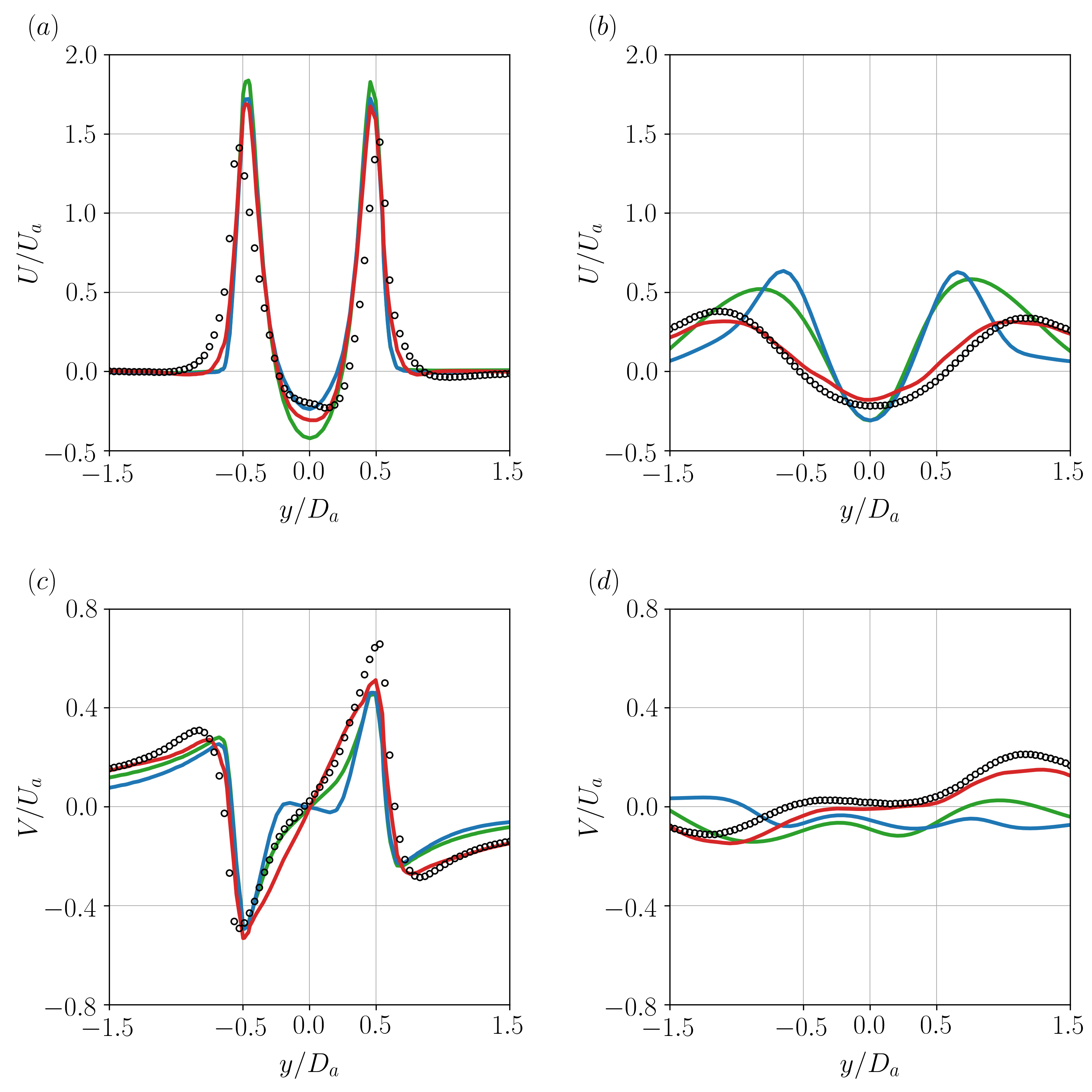}
	\caption{Impact of the turbulence model on the comparison with the PIV data. Mean velocities for the {\nog} model : mean axial velocity at (a) $x/{D_a}$ = 0; (b) $x/{D_a}$ = 2; mean radial velocity at (c) $x/{D_a}$ = 0; (d) $x/{D_a}$ = 2; {\protect\greenline} RKE; {\protect\blueline} SST; {\protect\redline} SBES; {\protect\blackcircle} PIV.} 
	\label{fig:Comparison}
\end{figure}

Initially we investigated the effect of the choice of the turbulence modelling approach. Figure \ref{fig:Comparison} shows a comparison of the time-averaged axial $U$ and radial $V$ velocity components predicted by the CFD modelling with the PIV data. The axial and radial coordinates are represented by $x$ and $y$, respectively. The velocity components have been normalised by the jet-exit mean velocity $U_a$, and the spatial coordinates by the jet-exit diameter $D_a$.  Comparisons are presented at two representative downstream lines, located just after the exit from the device and two diameters further downstream. It is evident that in all cases the SBES predictions are closer to the experimental data. In particular the realisable $k$-$\epsilon$ URANS models tends to over-predict the back-flow at the device outlet and the radial velocities distributions are much closer to the measured data for the SBES model. Given the importance of the prediction of the jet spreading rate, the use of the URANS models was discontinued.  

\subsection{Effect of the Grid}

\begin{figure}[h!]
	\centering
	\begin{subfigure}{1.0\textwidth}
	\centering
	\includegraphics[scale=0.4]{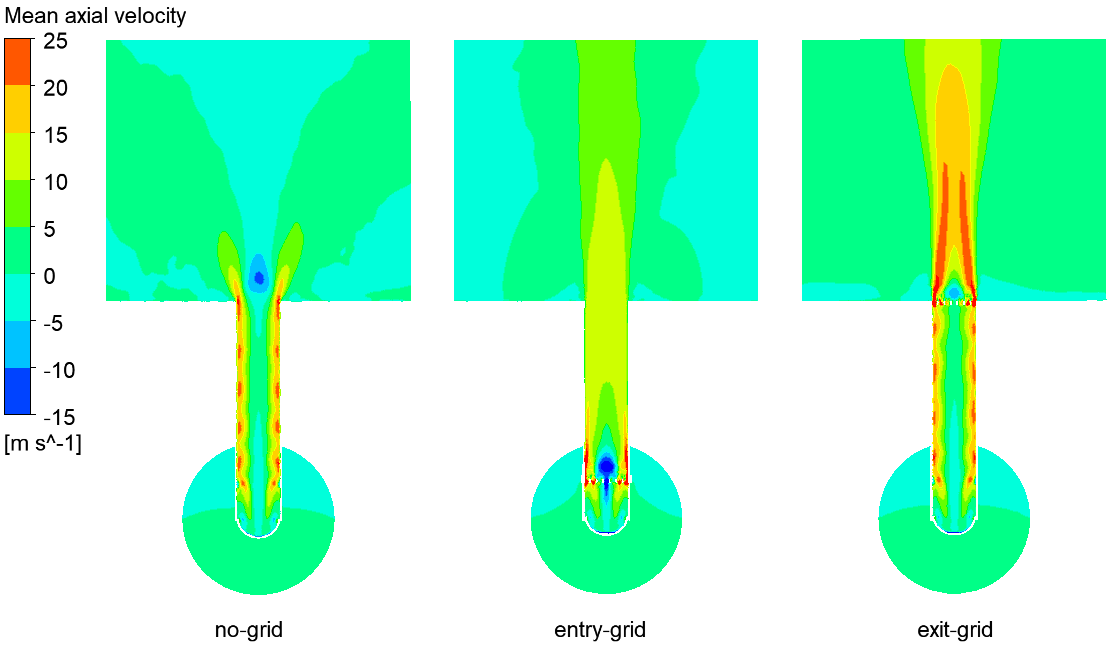} 
	\caption{}
	\end{subfigure} 
	\\
	\vspace{0.5em}
	\begin{subfigure}{1.0\textwidth}
	\centering
	\includegraphics[scale=0.4]{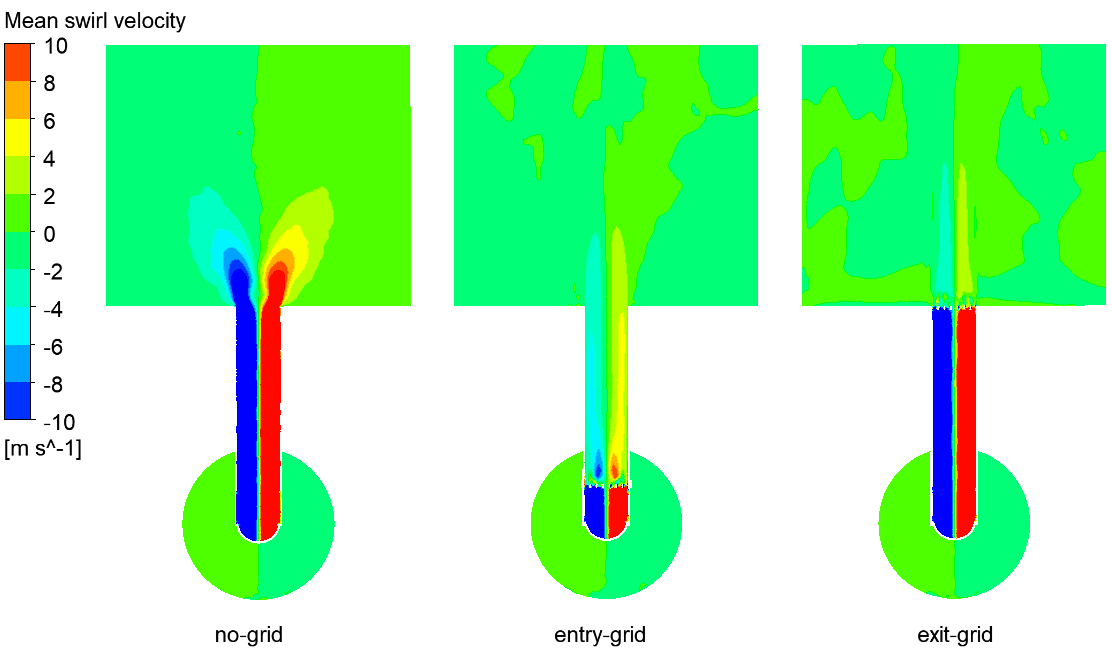} 
	\caption{}
	\end{subfigure} 
	\caption{Flow-field contour plots: (a) mean axial velocity ; (b) mean swirl velocity}
     \label{fig:GridComparison}
\end{figure}

The impact of the grid on the flow field is shown in Figure \ref{fig:GridComparison}, which presents the axial and swirl velocity components on a centre-plane. From Figure \ref{fig:GridComparison}(a) it is  apparent that the case with {\nog} shows a large vortex breakdown region at the exit of the device which leads to back-flow in the central region and as a consequence the wide dispersion of the axial flow. The {\eng} case shows much reduced jet spreading and the {\exg} case shows focusing of the high velocity jet generated by the grid towards the central axis. Both flow fields for devices with grids are potentially beneficial in that they are likely to focus particles along the centre of the jet.

The swirl velocities, given in Figure \ref{fig:GridComparison}(b), show the strong swirling flow exiting the device in the absence of a grid and that it is significantly reduced by the presence of the grid. In the {\eng} case the region of strong swirl is small and this may have an effect on particle de-agglomeration, whereas the {\exg} model shows strong swirl within the device being suppressed at the exit. 

\begin{figure}[h!]
	\centering
	\includegraphics[scale=0.4]{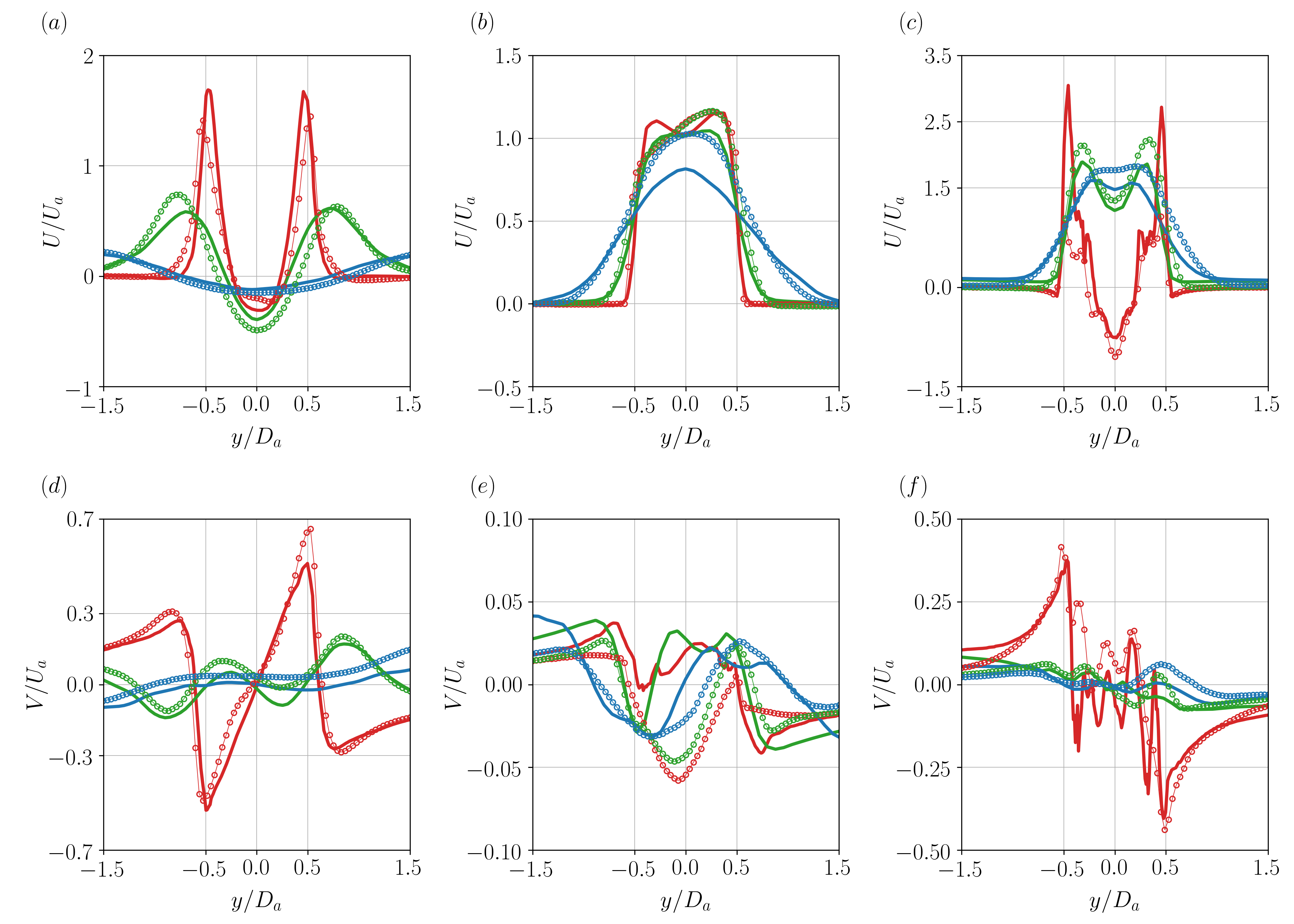}
	\caption{Mean axial and radial velocities for: (a) and (d) \nog; (b) and (e) \eng ; (c) and (f) {\exg} models; SBES: {\protect\redline} $x/{D_a}$ = 0, {\protect\greenline} $x/{D_a}$ = 1, {\protect\blueline} $x/{D_a}$ = 3; PIV: {\protect\redmarkline} $x/{D_a}$ = 0, {\protect\greenmarkline} $x/{D_a}$ = 1, {\protect\bluemarkline} $x/{D_a}$ = 3.} 
	\label{fig:Mean velocities}
\end{figure}

\newpage
Validation of the above flow fields was performed via comparison with detailed PIV data. Figure \ref{fig:Mean velocities} shows a comparison of the mean axial and radial velocity components with the PIV data. It is evident that in all cases there is good agreement between simulations and experiment. Mean axial velocities are well predicted with the worse agreement being a slight under-prediction of the central values at $x/D_a = 3$  for the {\eng} case. There are also some differences in the radial velocity in this case but the velocities are small and much less important in determining the flow field.  

\begin{figure}[h!]
	\centering
	\includegraphics[scale=0.4]{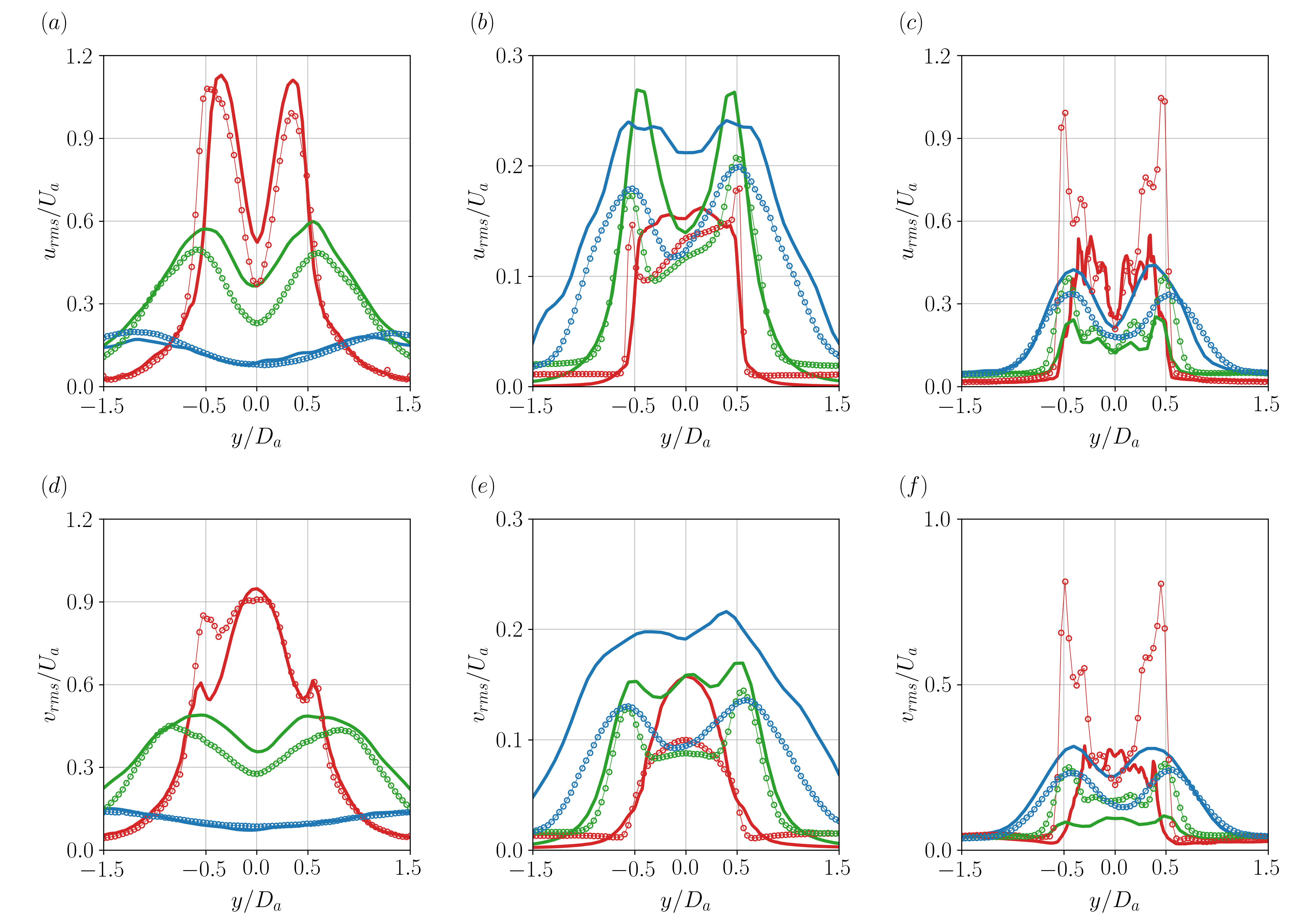}
	\caption{RMS axial and radial fluctuating velocities for: (a) and (d) \nog; (b) and (e) \eng ; (c) and (f) {\exg} models; SBES: {\protect\redline} $x/{D_a}$ = 0, {\protect\greenline} $x/{D_a}$ = 1, {\protect\blueline} $x/{D_a}$ = 3; PIV: {\protect\redmarkline} $x/{D_a}$ = 0, {\protect\greenmarkline} $x/{D_a}$ = 1, {\protect\bluemarkline} $x/{D_a}$ = 3.} 
	\label{fig:RMS velocities}
\end{figure}

\begin{figure}[h!]
	\centering
	\includegraphics[scale=0.38]{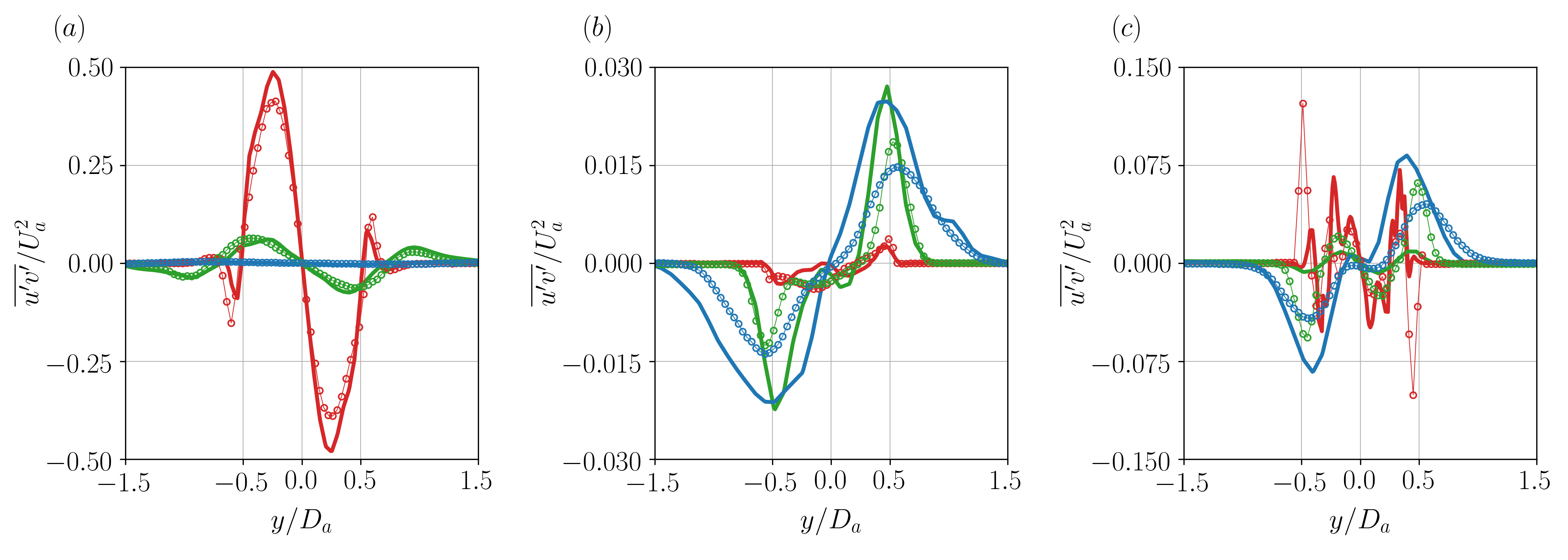}
	\caption{Reynolds shear-stress for: (a) \nog; (b) \eng ; (c) {\exg} models; SBES: {\protect\redline} $x/{D_a}$ = 0, {\protect\greenline} $x/{D_a}$ = 1, {\protect\blueline} $x/{D_a}$ = 3; PIV: {\protect\redmarkline} $x/{D_a}$ = 0, {\protect\greenmarkline} $x/{D_a}$ = 1, {\protect\bluemarkline} $x/{D_a}$ = 3.} 
	\label{fig:Reynolds stress}
\end{figure}

\newpage
Figures \ref{fig:RMS velocities} and \ref{fig:Reynolds stress} show the axial and radial velocity fluctuations and Reynolds stress comparisons with the PIV data. The best agreement is observed for the {\nog} case. However, whilst there are some deviations in the cases where grids are present, these are relatively small and are most pronounced close to the device in the {\exg} case. In this case, small deviations of measuring locations and fabrication tolerances would have the most pronounced effect. What is clear is that the CFD results correctly capture the magnitude and trends of these quantities in all cases, providing confidence for it to be used to investigate the entire flow field.

\subsection{Impact on the Pressure Drop}

\begin{figure}[h!]
	\centering
	\includegraphics[scale=0.35]{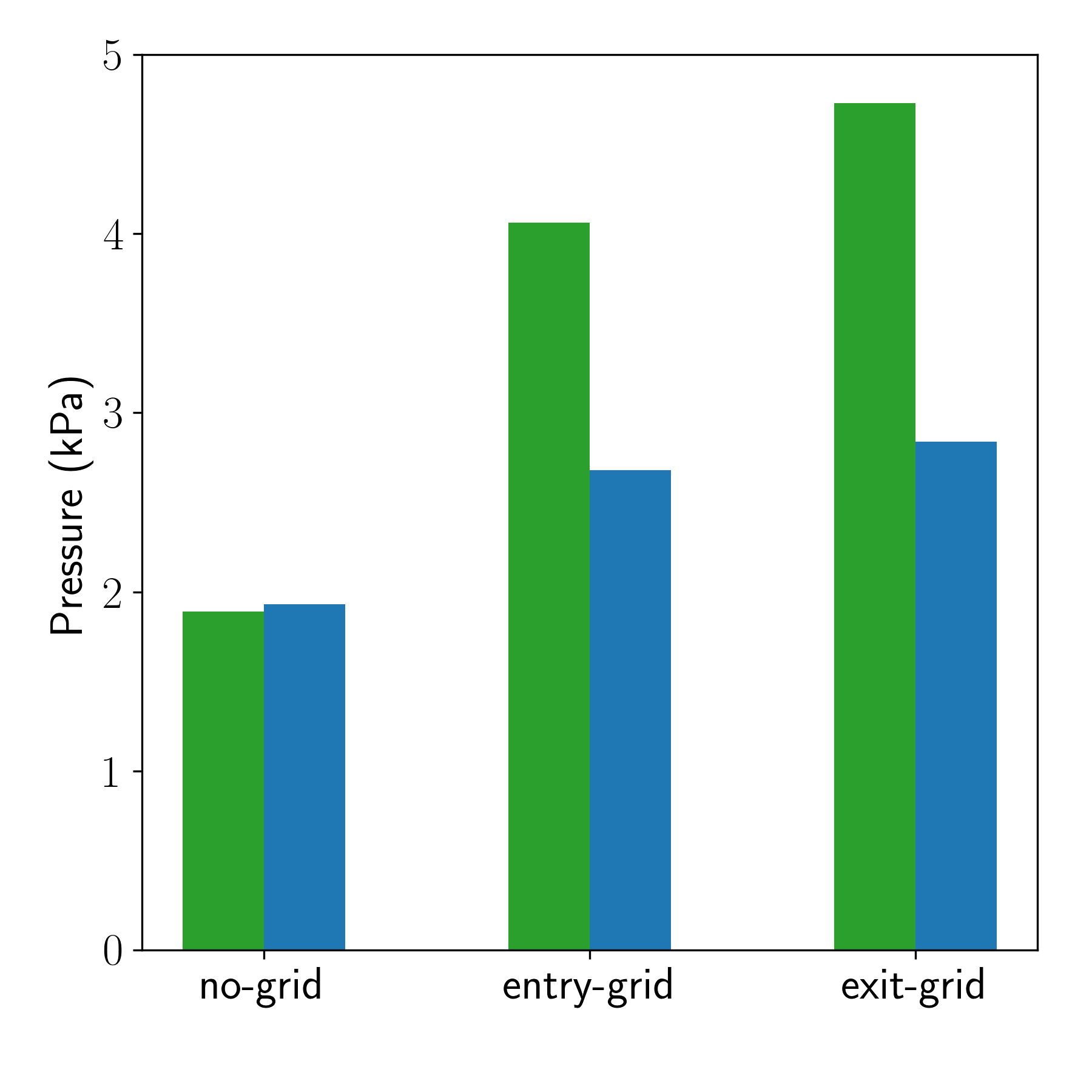}
	\caption{Pressure drop across the device models: {\protect\greenline} Measured; {\protect\blueline} CFD.} 
	\label{fig:Pressure drop}
\end{figure}

The measured pressure drop data are compared with the mean values obtained from the simulation in Figure \ref{fig:Pressure drop} for an air flow rate of \SI{60}{\litre\per\minute}. In the absence of a grid the values are very close, while the trend is correctly predicted, the value is under-predicted by about 35\%, for the two cases with a grid. The reason for this is unclear but is most likely related to small differences between the CAD geometry used to construct the CFD model and the 3D-printed physical device model,  and the surface roughness of the physical model.

\subsection{Influence on the Carrier Particles}

\begin{figure}[h!]
	\centering
	\begin{subfigure}{0.45\textwidth}
	\centering
	\includegraphics[scale=0.4]{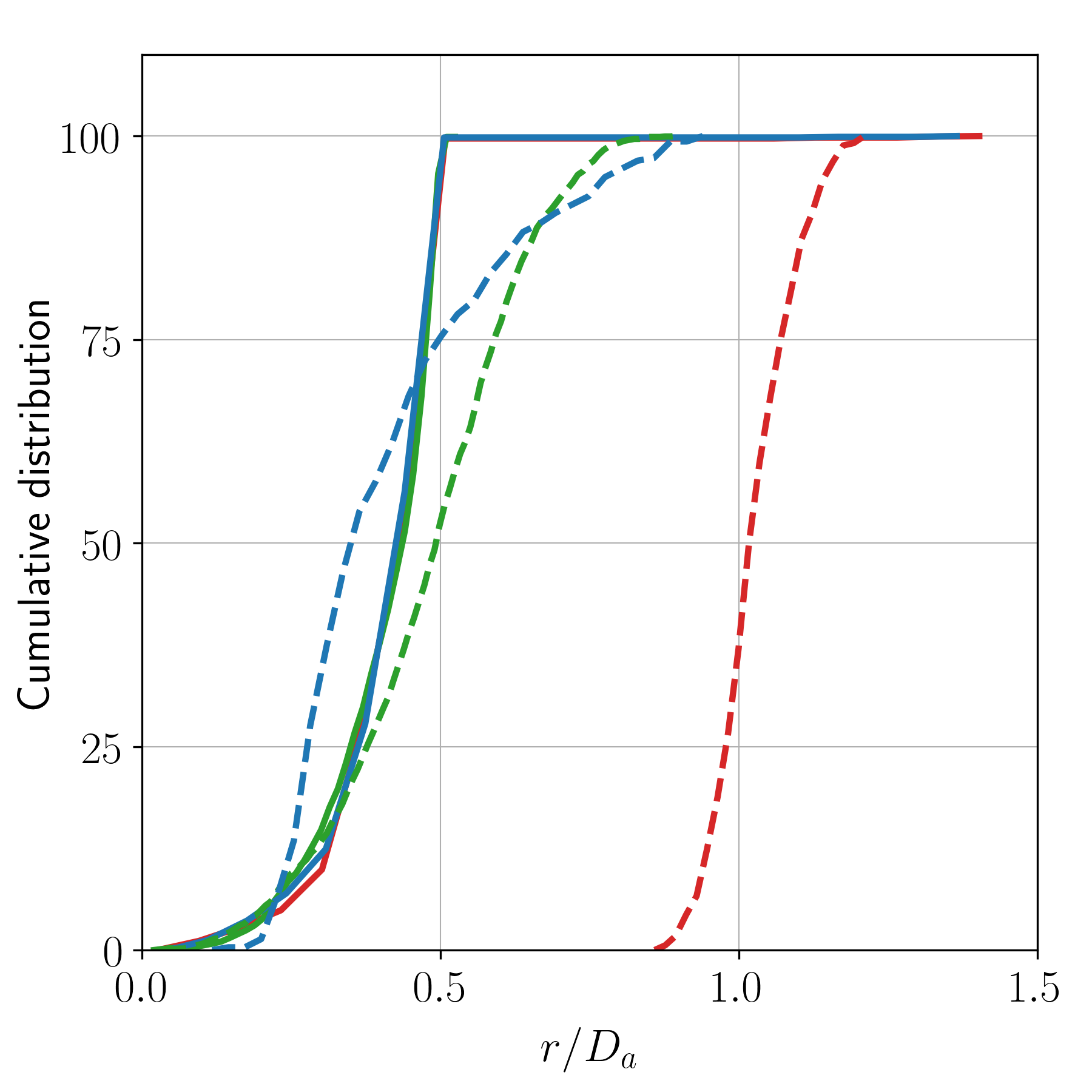} 
	\caption{}
	\end{subfigure} 
	\begin{subfigure}{0.45\textwidth}
	\centering
	\includegraphics[scale=0.4]{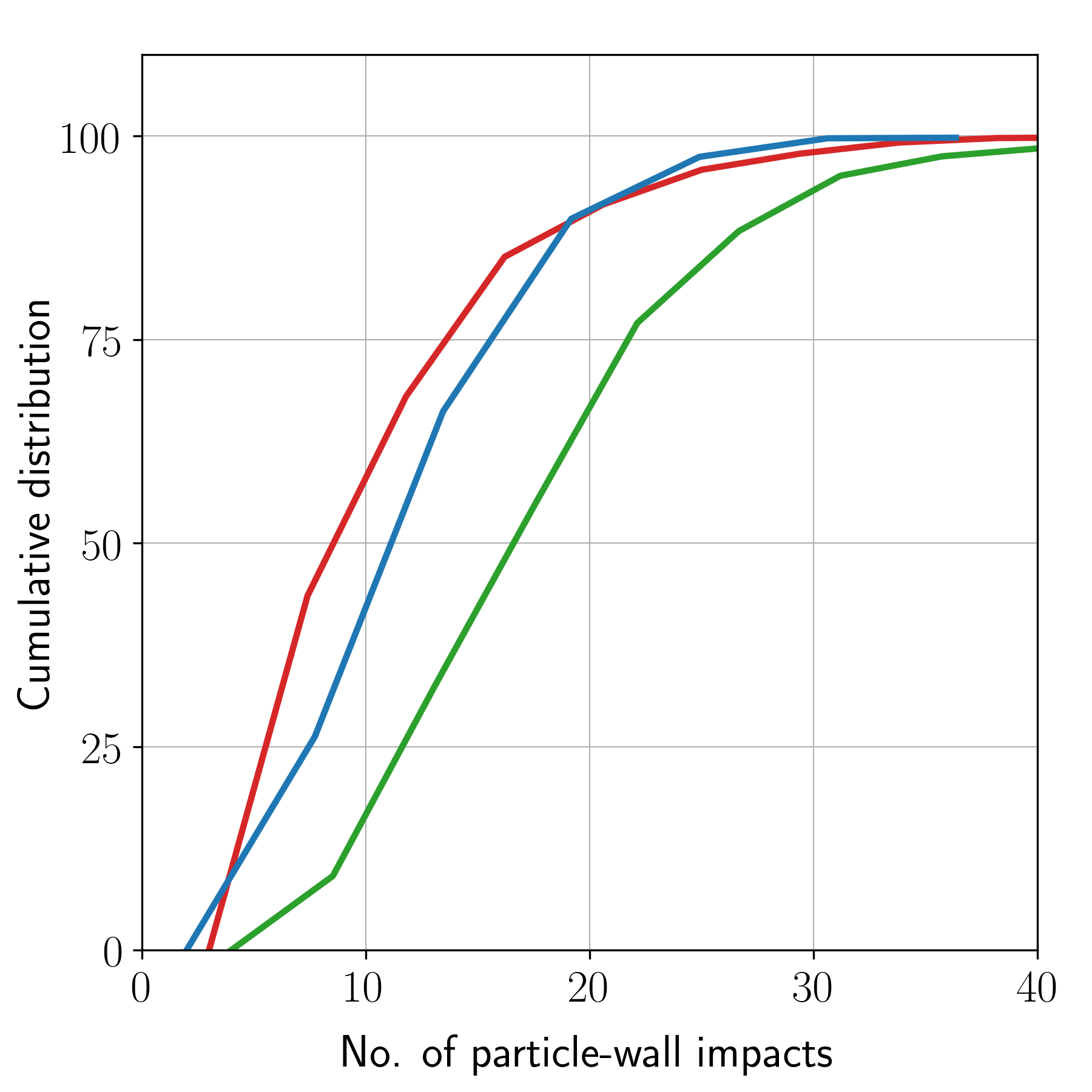} 
	\caption{}
	\end{subfigure}
	\\
	\begin{subfigure}{0.45\textwidth}
	\centering
	\includegraphics[scale=0.4]{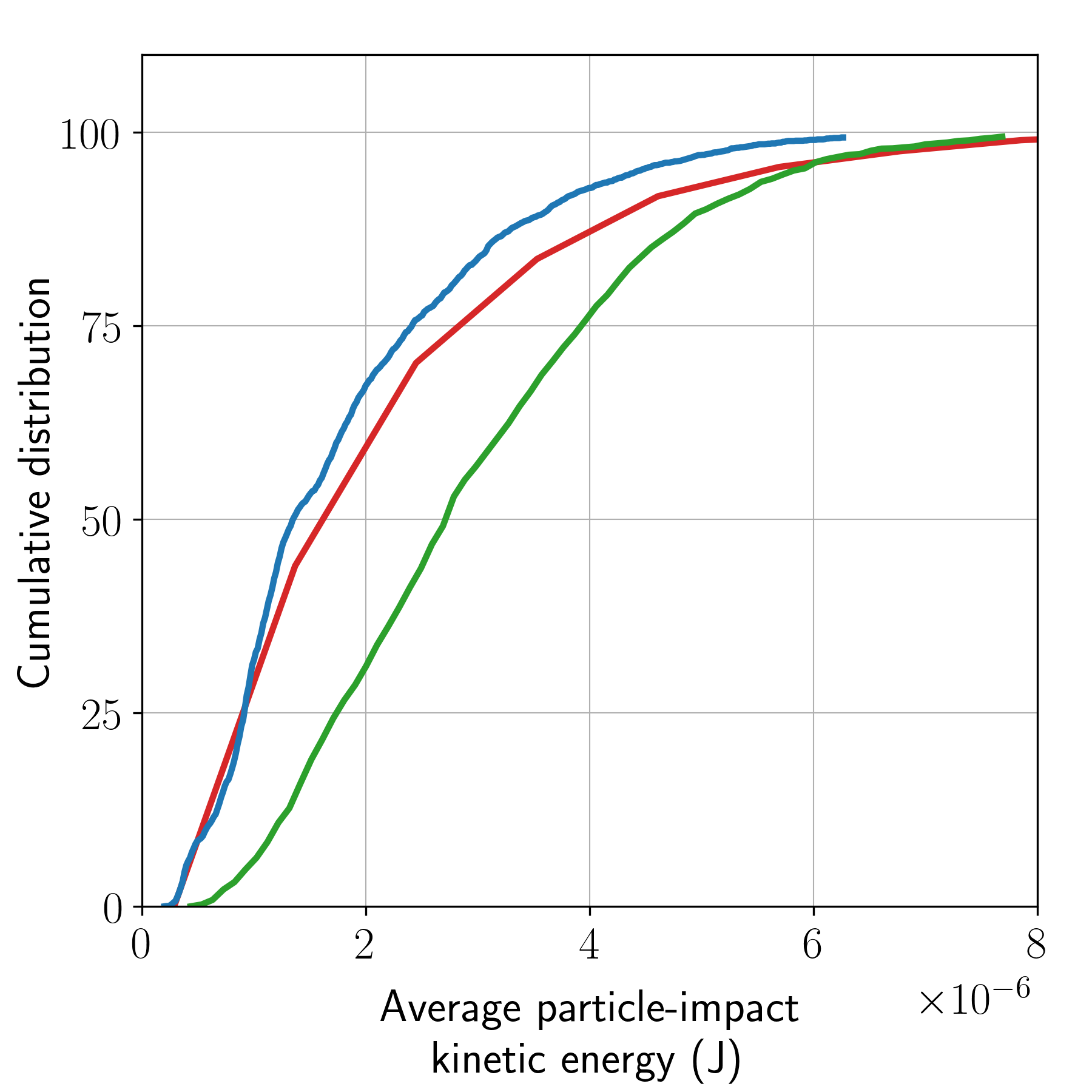} 
	\caption{}
	\end{subfigure}
	\caption{Cumulative distributions of particle variables: (a) Particle radial location, \protect\redline \nog, \protect\greenline \eng, \protect\blueline {\exg} at $x/{D_a}$ = 0, \protect\redbrokenline \nog, \protect\greenbrokenline \eng, \protect\bluebrokenline {\exg} at $x/{D_a}$ = 1 ; (b) Number of particle-wall impacts ; (c) Average particle-impact kinetic energy; \protect\redline \nog, \protect\greenline \eng, \protect\blueline \exg.}
	\label{fig:Particles}
\end{figure}

As discussed in Section \ref{particle_model}, carrier particles were released in the dosing cup of the device and their paths were tracked to collect data on spreading and wall impacts.  Figure \ref{fig:Particles}(a) shows the radial distribution of particles across the device at the exit and one jet-exit diameter downstream. For all three cases the exit distribution is very similar with particles clustered around the device wall. Even in the case with an {\eng} there is sufficient swirl to keep the particles at the wall. However, once they exit the device there is a very clear difference in behaviour. The particles in the {\nog} case have all moved in the radial direction by one jet-exit diameter and continue to move along that trajectory (data not shown). In the {\eng} case there is a small amount of outward spreading and in the {\exg} case there is spreading both inwards and outwards. The Stokes number for the particles is in the intermediate range ($\sim$0.3), so this behaviour is readily explained by the flow fields shown in Figure \ref{fig:GridComparison}, as once a particle leaves the inhaler it will tend to follow its initial trajectory while slowly responding to the influence of the flow.

Figure \ref{fig:Particles}(b) shows the average number of wall impacts per particle for the three cases. In terms of particle impacts, the best performing system is the {\eng} case, followed by the {\exg} and the worse is the case with {\nog}, with the median number of impacts in these cases being 16, 11 and 8, respectively. Clearly, the presence of a grid promotes particle-wall impacts but it is interesting that the {\eng} case has the best performance in this sense. The same trend is  present in the data for the mean particle impact energy in Figure \ref{fig:Particles}(c), with the median value for the {\eng} case being about twice that of the other two cases.

\newpage
Based on the above data, it is clear that a grid is important to reduce the particle spread and that the presence of a grid increases both the number of particle wall impacts and their energy. According to these predictions the {\eng} device should perform best.  

\begin{figure}[h!]
	\centering
	\includegraphics[scale=0.45]{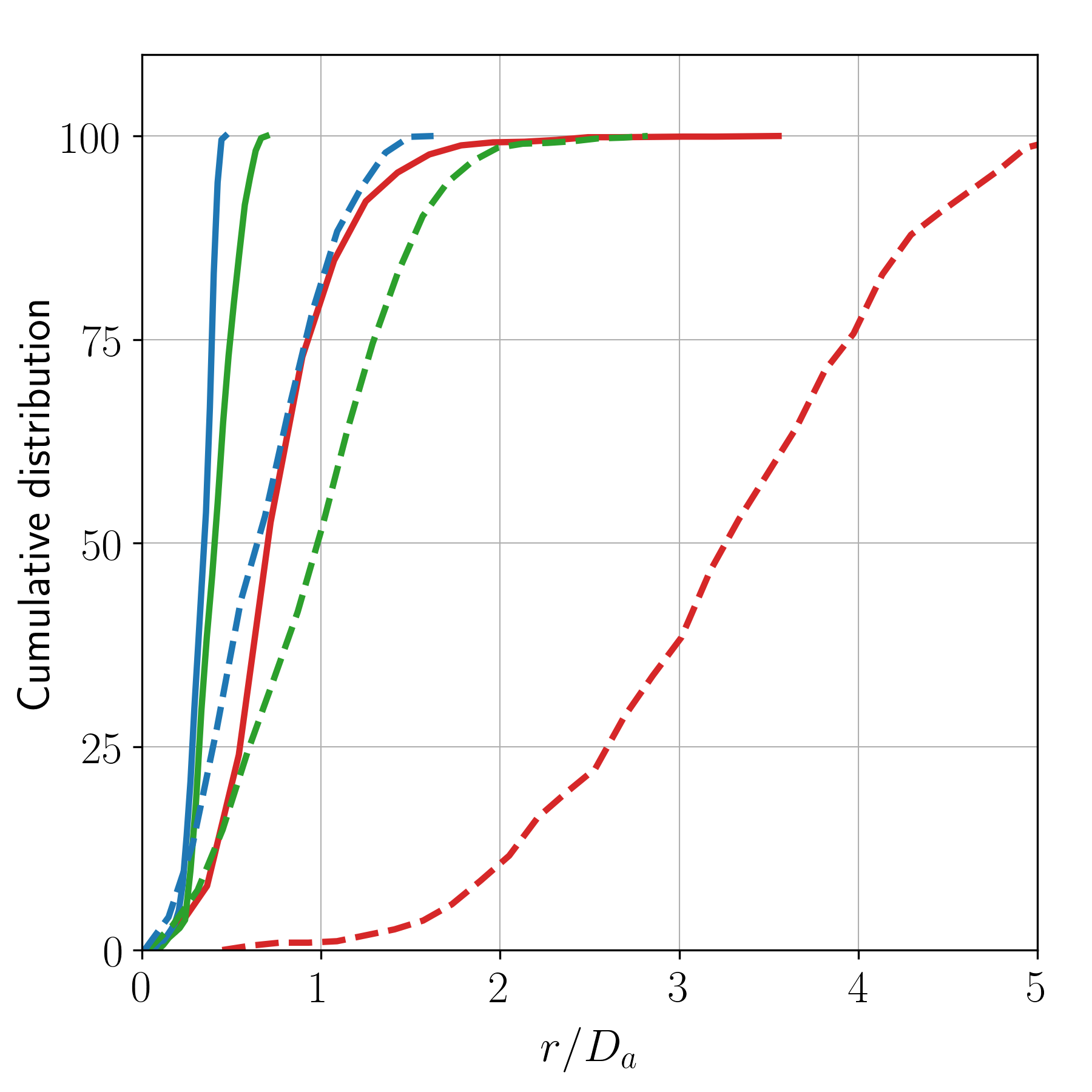}
	\caption{Cumulative distribution of radial location for the fine particles: \protect\redline \nog, \protect\greenline \eng, \protect\blueline {\exg} at $x/{D_a}$ = 1, \protect\redbrokenline \nog, \protect\greenbrokenline \eng, \protect\bluebrokenline {\exg} at $x/{D_a}$ = 5.} \label{fig:Fine particles}
\end{figure}

\subsection{Influence on the Drug Particles}

Figure \ref{fig:Fine particles} shows the spreading of the fine particles once they exit the device. At one jet-exit diameter downstream, the particles in the {\nog} case have already spread out around 1.5 jet diameters from the axis, whereas for the {\eng}  there is almost no spreading and there is a slight focusing effect in the {\exg} case. At 5 jet-exit diameters downstream, the fine particles are spread over 5 jet diameters in the {\nog} case, whereas the spreading is only 1.5 and 2 diameters for the {\exg} and {\eng}, respectively. Thus if reduced dispersion, and consequently less mouth-cavity deposition of the active ingredient is the aim, the {\exg} device is to be preferred based on these results.

\section{Discussion}

The objective of this paper was to perform CFD studies for a  number of inhaler designs and to confirm the results with experimental data in order to determine the utility of appropriate CFD simulations. Of course, this can only be done if the simulations results are of high quality and the models used are correctly applied. It takes experience and significant knowledge to do this correctly, so we have tried to outline the important questions to ask when setting up models and checking the results. For example, it was found that the common practice of applying boundary conditions at the device inlets leads to a non-physical influence on the flow in a very important part of the device. Similarly, the impact of using the correct turbulence modelling approach is highlighted. 
Whilst it is no surprise that these strongly swirling flows are time-dependent, it is clear that simply switching on transient flow, to change a RANS simulation to a URANS simulation, is not the correct approach. Doing this does indeed allow a transient simulation to made and the high residuals associated with an unconverged steady-state to be reduced but the URANS approach does not provide a physical description of the turbulence structure {\cite{Menter2018}}. What is now evident is that the earliest studies used relatively coarse computational meshes, lower order numerics and simpler turbulence models (without, for example, curvature correction terms) so that the flows often appeared much steadier than they do now because the swirl was artificially dissipated. The approach advocated here gives much more realistic turbulent flow fields meaning that both the swirl behaviour and the impact of the flow on particle transport are captured much more accurately. 

Validation against detailed PIV data has allowed the models to be assessed and it is clear that the SBES is a good approach, especially given the nature of the flow where there are significant regions of the flow domain occupied by attached boundary layers, which are known to be captured well using the SST model. The comparisons with PIV data presented herein provide very good validation of the modelling approach. This is important as CFD can then be used with confidence to explore the flow behaviour within the device itself, a region very difficult to access experimentally, and to screen ideas for new device designs. 

 The impact of the grid on mouth-cavity deposition is well captured in the simulations as the results conform with the {\invit} results showing that there was a significant difference between the devices, with most deposition in the {\nog} case and least in {\exg} case. The {\invit} studies showed that more drug remained in the device for the {\exg} case, a parameter that was not assessed in this model. Moreover, the fine particle fraction (FPF) in the {\invit} study was similar amongst the devices, with values of 52.83\% \begin{math} \pm \end{math} 3.45, 53.05\% \begin{math} \pm \end{math} 7.17 and 56.25\% \begin{math} \pm \end{math} 4.54 for the {\nog}, {\eng} and {\exg}, respectively. From the CFD results presented here, the presence of the grid led to a higher mean number of impacts and increased impact kinetic energy of the particles, which is expected to translate into greater drug detachment from the carrier particles. Although there was a numerical increase in FPF, the increased number of particle-wall impacts observed in the CFD did not lead to a significant increase in FPF, as shown in a previous study \cite{dosReis2020}. During aerosolization, drug detachment from the carrier is thought to derive from both particle-wall and particle-particle collisions. From CFD results, the {\eng} case was predicted to have a better performance due to its greater de-agglomeration potential resulting from the higher number of particle-wall impacts. However, particle-particle collisions were not modelled in this study, which is the likely explanation for the differences observed between CFD and {\invit} results.

\section{Conclusions}

This paper has shown that provided the correct modelling choices are made and the simulations are executed with the appropriate care and knowledge, CFD can provide significant insights into DPI performance. Simulations using the Stress Blended Eddy Simulation (SBES) approach are well suited for this task, which is supported by the very good agreement with the PIV data. This turbulence modelling choice is important as it allows the transient nature of the flow and the significant turbulence generation by highly swirling flows to be captured. This has a follow-on effect on the dispersion of the fine particles that have low Stokes numbers and follow the turbulent eddies. This work shows that it is possible to improve upon the use of RANS or URANS significantly without going to a full LES simulation. In particular, the proposed approach uses the optimal turbulence modelling approach in each zone: RANS in attached boundary layers at the walls and LES in the regions of separated flow and wakes. Use of pure LES is not practical as it requires locally refined meshes in all three dimensions at the wall if the boundary layer is to be captured correctly.

The simulations capture important experimental observations of the reduction in radial spreading of the flow and fine particles due to the presence of a grid, with the {\exg} geometry performing best, in line with the reduced mouth-cavity deposition observed in the experiments.
Keeping in mind that the experiments did not use a throat geometry and the simulations did not model all aspects of the particle behaviour, specifically particle-particle interactions and particle detachment the adopted CFD approach captured the dispersion data quite well. 

\section*{Acknowledgments}
The research was supported by the Australian Research Council. The authors acknowledge the University of Sydney for providing High Performance Computing resources that have greatly contributed to the research results reported here (http://sydney.edu.au/research{\textunderscore}support). The research was also benefited from computational resources provided through the NCMAS, supported by the Australian Government. The computational facilities supporting this project included the Multi-modal Australian ScienceS Imaging and Visualisation Environment (MASSIVE) at Monash.

\clearpage
\bibliography{References-paper2-updated} 
\end{document}